\documentclass{emulateapj}

\usepackage{graphics}
\usepackage{epsfig}
\usepackage{natbib}
\shorttitle{High-velocity Hydrogen in SN 1987A}
\shortauthors{France et al.}
\begin{document}

\title{Mapping  High-velocity H-$\alpha$ and Lyman-$\alpha$ Emission from Supernova 1987A} 


\author{
Kevin France\altaffilmark{1,2},
Richard McCray\altaffilmark{3}, 
Claes Fransson\altaffilmark{4},
Josefin Larsson\altaffilmark{5}, 
Kari A. Frank\altaffilmark{6},
David N. Burrows\altaffilmark{6},
Peter Challis\altaffilmark{7},
Robert P. Kirshner\altaffilmark{7},
Roger A. Chevalier\altaffilmark{8}, 
Peter Garnavich\altaffilmark{9}, 
Kevin Heng\altaffilmark{10},
Stephen S. Lawrence\altaffilmark{11},
Peter Lundqvist\altaffilmark{4},  
Nathan Smith\altaffilmark{12}, 
George Sonneborn\altaffilmark{13}
}


\altaffiltext{1}{Laboratory for Atmospheric and Space Physics, University of Colorado, 392 UCB, Boulder, CO 80309;  kevin.france@colorado.edu}
\altaffiltext{2}{Center for Astrophysics and Space Astronomy, University of Colorado, 389 UCB, Boulder, CO 80309}
\altaffiltext{3}{Department of Astronomy, University of California, Berkeley, CA 94720-3411}
\altaffiltext{4}{Department of Astronomy, The Oskar Klein Centre, Stockholm University, SE-106 91 Stockholm, Sweden}
\altaffiltext{5}{KTH, Department of Physics, and the Oskar Klein Centre, AlbaNova, SE-106 91 Stockholm, Sweden}

\altaffiltext{6}{Department of Astronomy and Astrophysics, The Pennsylvania State University, 525 Davey Lab, University Park, PA 16802 USA}
\altaffiltext{7}{Harvard-Smithsonian Center for Astrophysics, 60 Garden Street, MS-19, Cambridge, MA 02138, USA}
\altaffiltext{8}{Department of Astronomy, University of Virginia, P.O. Box 400325, Charlottesville, VA 22904-4325, USA }
\altaffiltext{9}{225 Nieuwland Science, University of Notre Dame, Notre Dame, IN 46556-5670, USA}
\altaffiltext{10}{University of Bern, Center for Space and Habitability, Sidlerstrasse 5, CH-3012 Bern, Switzerland }
\altaffiltext{11}{Department of Physics and Astronomy, Hofstra University, Hempstead, NY 11549, USA}
\altaffiltext{12}{Steward Observatory, University of Arizona, 933 North Cherry Avenue, Tucson, AZ 85721, USA}
\altaffiltext{13}{NASA Goddard Space Flight Center, Code 665, Greenbelt, MD 20771, USA}

%


\begin{abstract}

We present new {\it Hubble Space Telescope} images of high-velocity H-$\alpha$ and Lyman-$\alpha$ emission in the outer debris of SN~1987A.   The H-$\alpha$ images are dominated by emission from hydrogen atoms crossing the reverse shock.  For the first time we observe emission from the reverse shock surface well above and below the equatorial ring, suggesting a bipolar or conical structure perpendicular to the ring plane.  Using the H$\alpha$ imaging, we measure the mass flux of hydrogen atoms crossing the reverse shock front, in the velocity intervals ($-$7,500~$<$~$V_{obs}$~$<$~$-$2,800 km s$^{-1}$) and (1,000~$<$~$V_{obs}$~$<$~7,500 km s$^{-1}$), $\dot{M_{H}}$ = 1.2~$\times$~10$^{-3}$ M$_{\odot}$ yr$^{-1}$.  We also present the first Lyman-$\alpha$ imaging of the whole remnant and new $Chandra$ X-ray observations.  Comparing the spatial distribution of the Lyman-$\alpha$ and X-ray emission, we observe that the majority of the high-velocity Lyman-$\alpha$ emission originates interior to the equatorial ring.  The observed Lyman-$\alpha$/H-$\alpha$ photon ratio, $\langle$$R(L\alpha / H\alpha)$$\rangle$ $\approx$~17, is significantly higher than the theoretically predicted ratio of $\approx$ 5 for neutral atoms crossing the reverse shock front.  We attribute this excess to Lyman-$\alpha$ emission produced by X-ray heating of the outer debris.   The spatial orientation of the Lyman-$\alpha$ and X-ray emission suggests that 
X-ray heating of the outer debris is the dominant Lyman-$\alpha$ production mechanism in SN 1987A at this phase in its evolution.  

\end{abstract}

\clearpage

\section{Introduction}

The reverse shock (RS) in SN 1987A is the surface where the freely expanding debris is suddenly decelerated as it encounters circumstellar matter.
Hydrogen atoms crossing the shock are excited by collisions in the shocked plasma and emit H-$\alpha$ and Lyman-$\alpha$ photons having Doppler shifts corresponding to the projected velocity of the freely expanding debris immediately inside the shock.  This emission was predicted by~\citet{borkowski97} and subsequently observed and interpreted by~\citet{sonneborn98} and~\citet{michael98}.\nocite{sonneborn98,michael03}  

The freely expanding supernova debris obeys a ``Hubble's Law'', i.e., the Doppler velocity is given by $V_{obs}$ = $z$/$t$ , where $z$ is the projected distance along the line of sight measured relative to the center of the explosion and $t$ is the time since the supernova explosion ($\approx$~27.5 years in mid-2014).  Thus, surfaces of constant Doppler shift are planar sections of the debris.  The fluxes of H-$\alpha$ and Lyman-$\alpha$ photons emitted at the RS are directly proportional to the fluence of hydrogen atoms across the shock.  

These facts create a unique opportunity to map the three-dimensional structure of the RS in SN~1987A and provide an empirical framework for studies of the shock interaction.  This opportunity has been exploited in observations with the Space Telescope Imaging Spectrograph (STIS).  In the STIS observations, the H-$\alpha$ emission from the RS was clearly visible near the circumstellar equatorial ring (ER), where the shock is strongest~\citep{michael98,michael03,heng06,france10c}.  The ring is inclined $\approx$~43\arcdeg, north being the near side (Panagia et al. 1991; Plait et al. 1995; Sugerman et al. 2005).  Michael et al. (2003) modeled the RS surface, concluding that the observed emission was confined to within $\sim$~$\pm$~30\arcdeg\ of the ring plane, similar to the opening angle of the ring observed at radio wavelengths~\citep{ng13}.  However, the STIS exposures were not deep enough to detect H-$\alpha$ emission at high latitudes (i.e., out of the ER plane), where the fluence of atoms is much smaller than near the equator.  Moreover, none of the spectroscopic observations included enough STIS slit locations to map the entire shock surface.\nocite{panagia91,sugerman05,plait95}

In this Letter, we describe the results of an observing campaign to map the high-velocity H-$\alpha$ and Lyman-$\alpha$ emission using filters that transmit the blue- and red-shifted emission from the debris while suppressing the bright emission from the ER. In Sections 2 and 3, the new $HST$ observations are described; for the first time we detect emission from the shock above and below the ring plane.  In Section 4, we compare H-$\alpha$, Lyman-$\alpha$, and X-ray observations, arguing that the majority of the Lyman-$\alpha$ flux is driven by X-ray heating of the outer ejecta, as proposed in our earlier work (e.g., Larsson et al. 2011; France et al. 2011; Fransson et al. 2013).\nocite{larsson11,france11,fransson13}


\begin{figure}
\begin{center}
\epsfig{figure=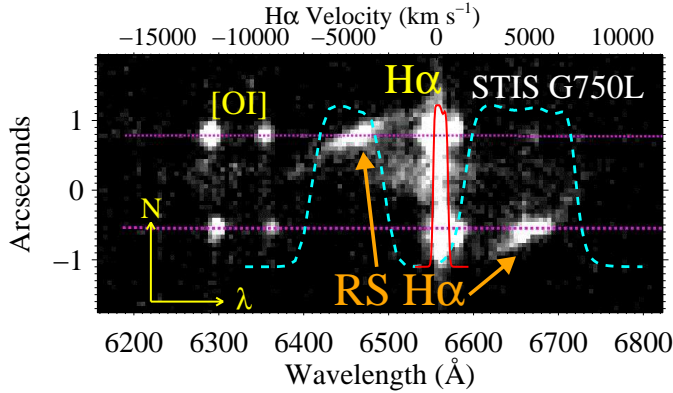,width=3.7in,angle=00}
\vspace{-0.1in}
\caption{
\label{cosovly} 
20 August 2014 STIS G750L spectra with WFC3 filter curves: dashed cyan lines represent F645N and F665N, while the solid red line represents the F656N filter.  
The F658N filter is shifted about 2 nm to the red of F656N and is not shown here.  
Narrow emission lines from the ER are labeled in yellow and the broad RS arcs are labeled in orange.  The dotted purple line shows the position of the equatorial ring.  These spectra have an effective slit width of 0.5\arcsec\ in the East-West direction of the images shown in Figure 2~--~4. }
\end{center}
\end{figure}

\section{$HST$ Observations and Image Reduction}

We obtained images of SN~1987A with the WFC3 camera in four filters to isolate the high-velocity H-$\alpha$:  blue high-velocity H-$\alpha$ (``H-$\alpha$ Blue'', F645N), red high-velocity H-$\alpha$ (``H-$\alpha$ Red'', F665N), and two narrow-band filters centered on shocked emission from the ER and hotspots (narrow H-$\alpha$, F656N and [\ion{N}{2}] $\lambda$~6583~\AA, F658N).   The WFC3 images each have exposure times of 2880 seconds and were acquired on 21 June 2014.   Figure 1 shows the band passes of these filters superposed on our 2014 STIS spectrum of SN~1987A (see also Larsson et al.~--~{\it in prep.}).  \nocite{fransson15,larsson15}

The ``H-$\alpha$ Blue'' and ``H-$\alpha$ Red'' images include a contribution from emission lines originating on the ER (see Fransson et al.~--~{\it in prep.} for a discussion of the ER emission). The narrow band images are used to subtract these contributions to the high velocity material images (Figure 2).  The F656N image provided the best template for the north-dominated high-velocity ``H-$\alpha$ Blue'' image.  The F658N filter contains a significant contribution from redshifted H-$\alpha$ on the southern side of the remnant, therefore this filter provided the best template for the south-dominated high-velocity ``H-$\alpha$ Red'' image.  
In Figure 2 (right), the ER appears black; however this is an artifact of the intentional oversubtraction and stretch of the images for display.  For the quantitative analysis in Section 4, a smaller ER subtraction is applied; the emission is nearly continuous across the north and south ER region.   

We observed the Lyman-$\alpha$ emission from SN 1987A with the ACS/SBC F122M filter for three orbits (9200 seconds; Figures 3 and 4).  The F122M filter bandpass is shown on an $HST$-COS spectrum~\citep{france11} in Figure 3.  In this case, resonant scattering by the neutral ISM of the Milky Way and Large Magellanic Cloud acts as a natural blocking filter for low-velocity emission from the ER and no additional subtraction is necessary.  A UV-bright B star (``Star 3'') appears to the southeast in all imaging bands, providing a common reference point.   We complement the $HST$ imaging with a recent (2014 March 19, day 9885) 0.3~--~8.0 keV image from $Chandra$~(Frank et al.~--~{\it in prep.}).

\begin{figure*}
\begin{center} 
\epsfig{figure=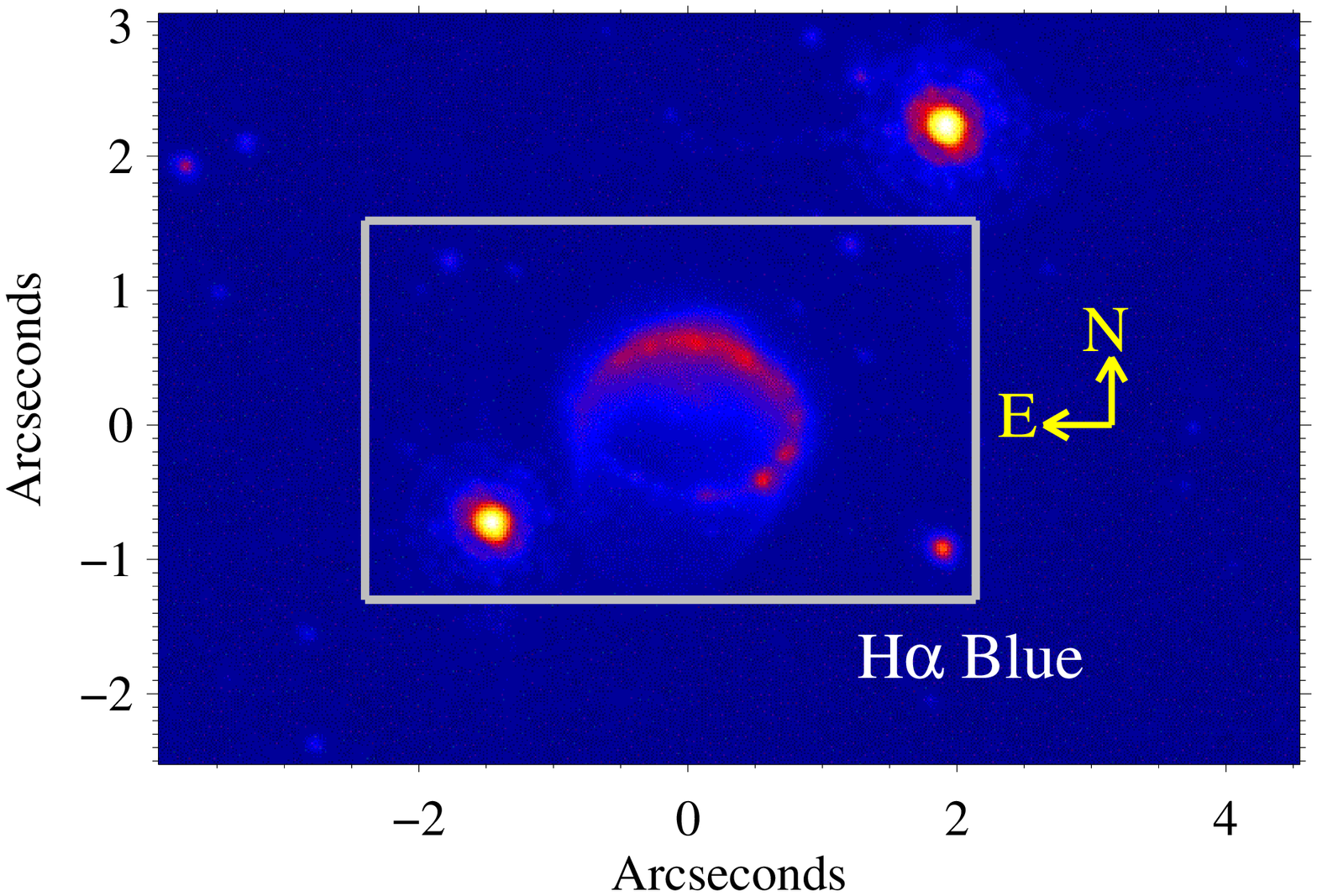,width=3.3in,angle=00} \hspace{-0.2in}
\epsfig{figure=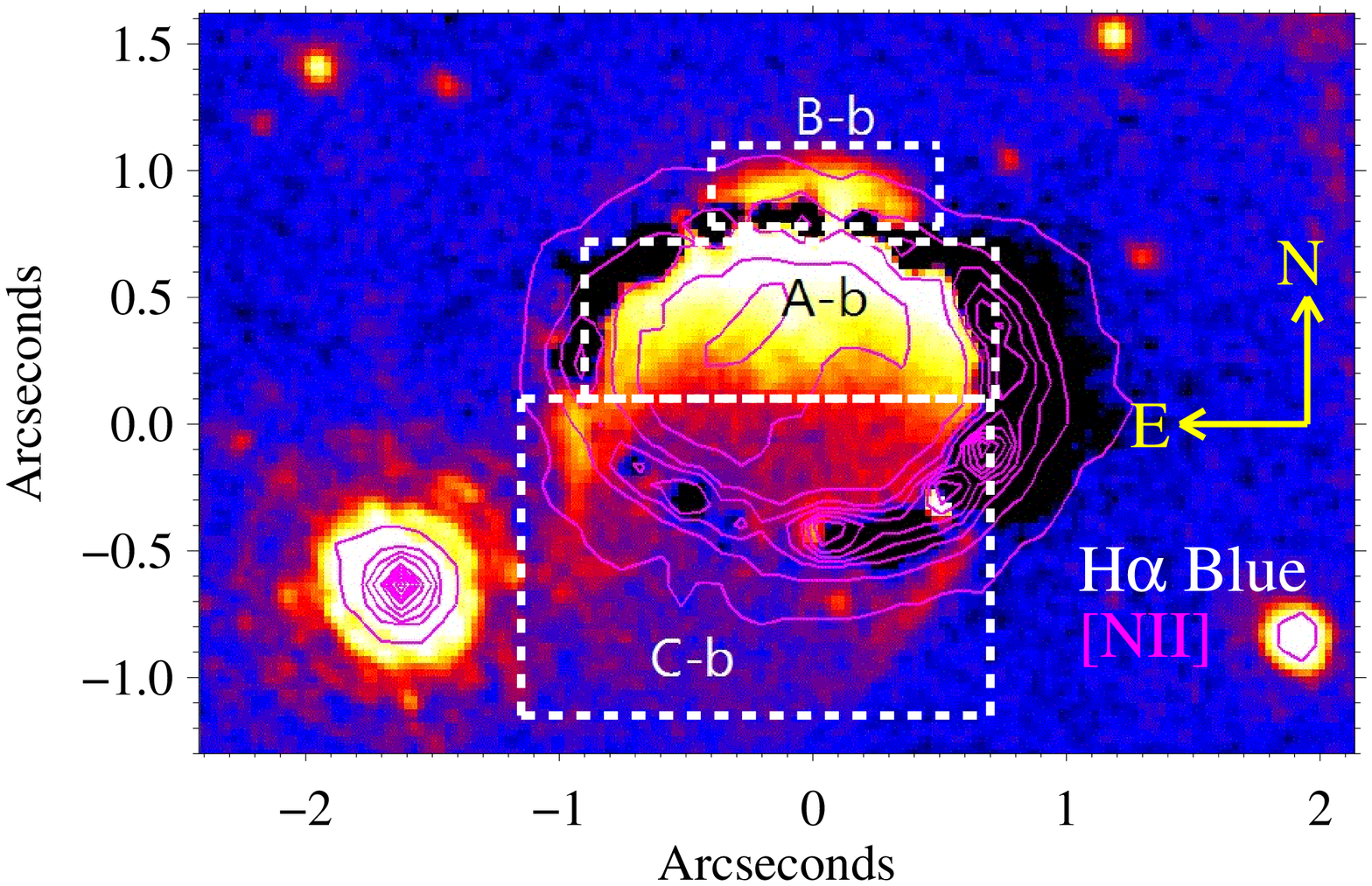,width=3.3in,angle=00}
\epsfig{figure=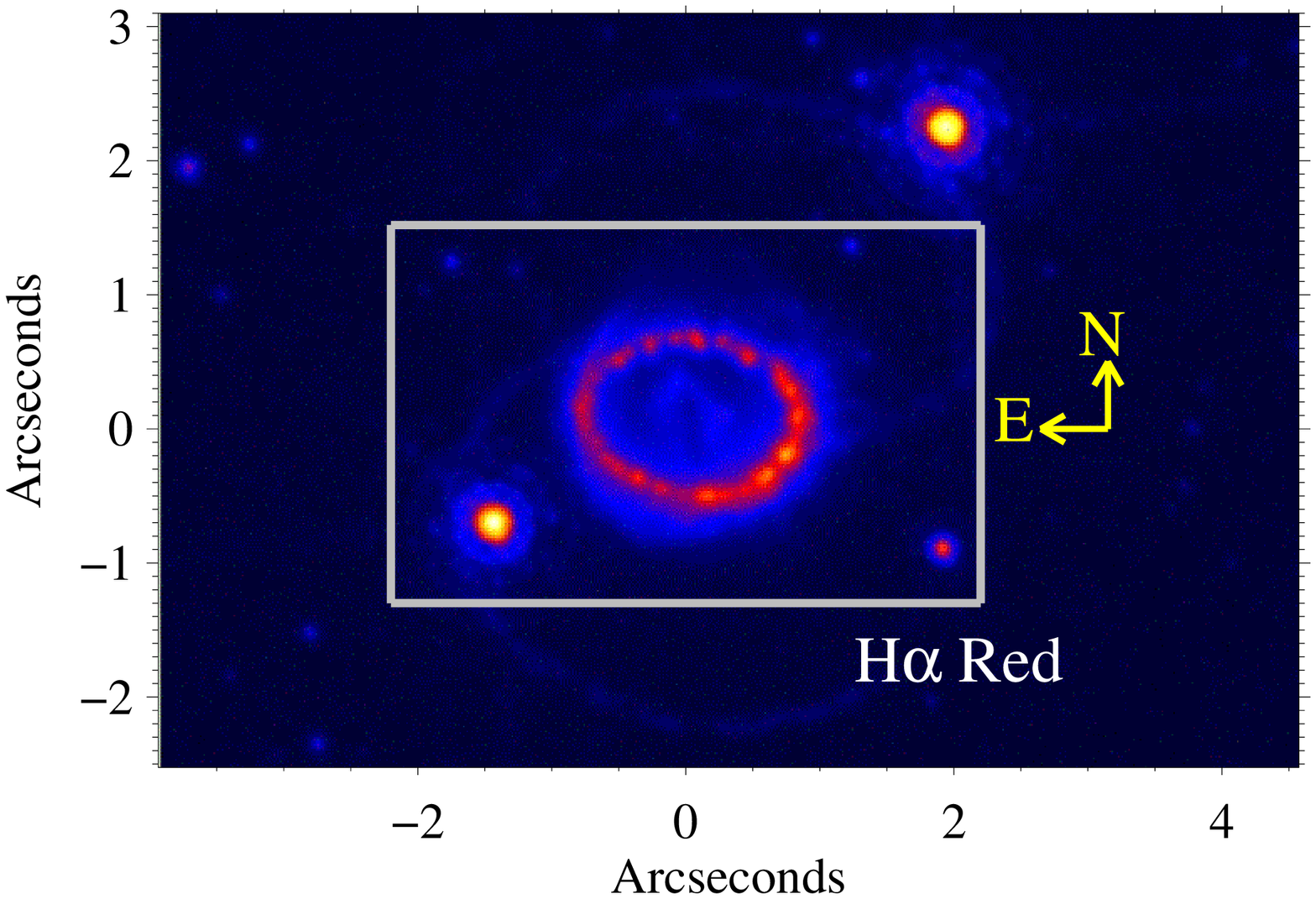,width=3.3in,angle=00} \hspace{-0.22in}
\epsfig{figure=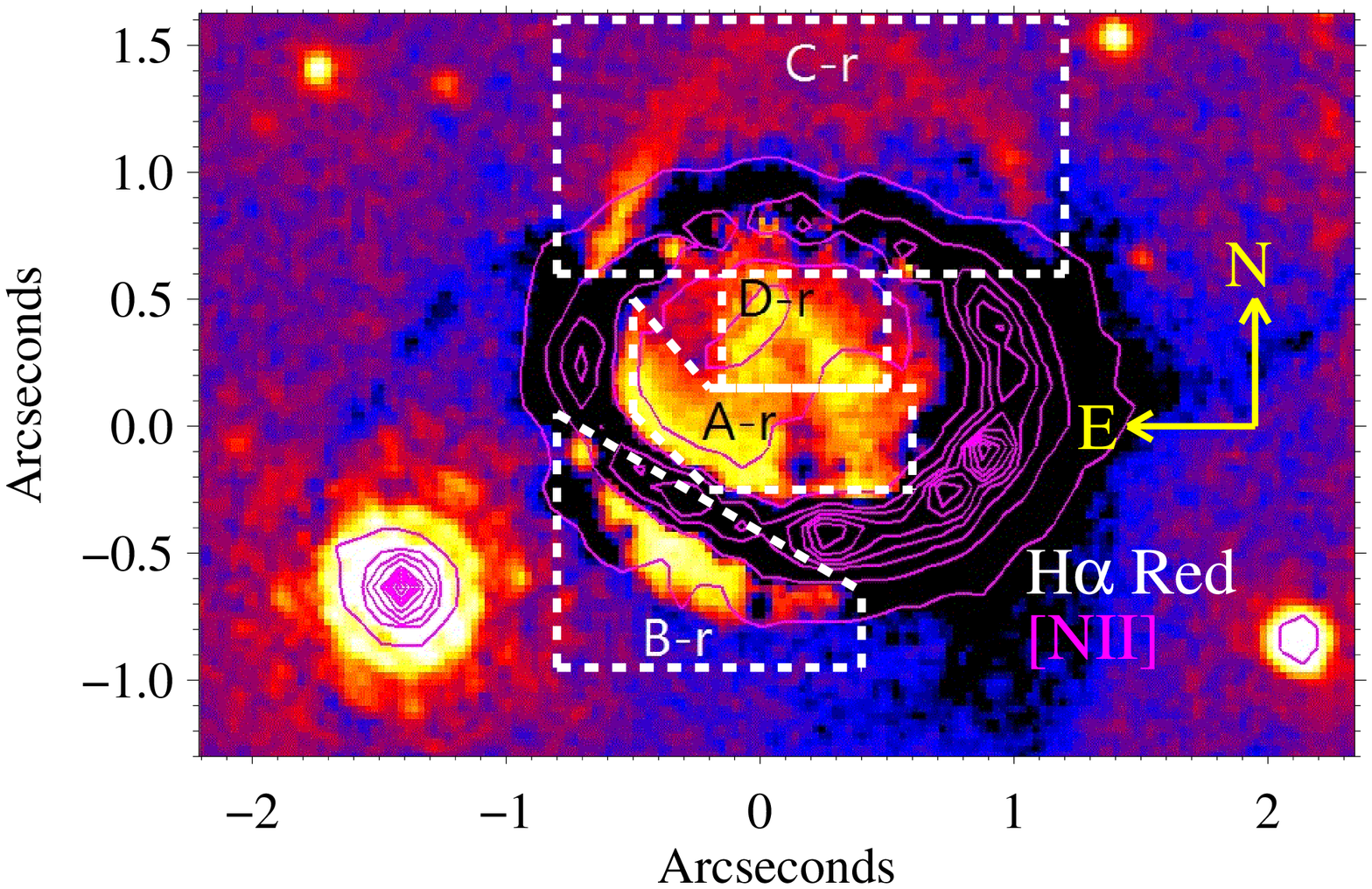,width=3.3in,angle=00}
\vspace{+0.05in}
\caption{
\label{cosovly} The left column shows the original F645N (``H-$\alpha$ Blue'') and F665N (``H-$\alpha$ Red'') images (21 June 2014), with the blow up region outlined in gray.  The right column shows an enlarged version after subtracting emission from the equatorial ring.  The [NII] contours are shown in magenta to identify the location of the ER.  The regions of interest are approximately outlined in dashed white lines.    
 }
\end{center}
\end{figure*}

\section{High-velocity H-$\alpha$ Imaging} 

The ``H-$\alpha$ Blue'' filter samples H-$\alpha$ emission with Doppler shifts  $-$7500 $<$ $V_{obs}$ $<$ $-$ 2800 km s$^{-1}$, while the ``H-$\alpha$ Red'' filter samples +1000 $<$ $V_{obs}$ $<$ +7500 km s$^{-1}$ (Figure 2).  
Regions of interest are labeled A, B, C, and D; ``-b'' and ``-r'' denote blue and red H-$\alpha$ velocities, respectively.  Emission from the A-b and A-r regions are produced by H atoms passing through the RS near the north and south side of the ER, respectively.
The bright H-$\alpha$ emission features (A-b and A-r) from the RS have been partially mapped in 3 dimensions by STIS observations (e.g., Michael et al. 2003; Heng et al. 2006; France et al. 2010).


In the ``H-$\alpha$ Blue'' image (Figure 2, top) a bright crescent-shaped feature covers the north side of the ER (as defined by the [\ion{N}{2}] contours), labeled B-b.  Given the 43\arcdeg\ inclination of the ring, the B-b region must originate from significantly above the ring plane for it to be visible in this configuration.    This image also shows faint streaks that cross the ring extending towards the south (C-b).  The features labeled A-r, B-r, and C-r are analogous to the blue side, but reflected about the major axis of the ER.  The additional feature labeled D-r is probably emission from the internal debris, which is expanding with radial velocities ranging up to $\sim$~3500 km s$^{-1}$~\citep{mccray93,larsson13}.

Combining the ``H-$\alpha$ Blue'' and ``H-$\alpha$ Red'' images, we measure the spatial extent of the high-velocity H-$\alpha$ regions (positional uncertainties $\lesssim$~0.05\arcsec).  The high-velocity material interior to the ring (A regions) has north-south and east-west diameters of 0.94\arcsec\ $\times$ 1.31\arcsec, which when including the inclination translates to physical dimensions of (9.9~$\times$~9.8)~$\times$~10$^{17}$ cm (assuming $d$~=~50 kpc).  This suggests a symmetric, radial flow with a maximum radial velocity of 5.7~$\times$~10$^{3}$ km s$^{-1}$ (at 27.5 years since the supernova explosion).  The B regions on the blue and red side have maximum angular distances from the center of the remnant of 0.82\arcsec\ and 0.98\arcsec, respectively.  These correspond to maximum de-projected distances and radial velocities of 8.6~$\times$~10$^{17}$ cm and  10.0~$\times$~10$^{3}$ km s$^{-1}$ for the north/blue and  10.4~$\times$~10$^{17}$ cm and  11.9~$\times$~10$^{3}$ km s$^{-1}$ for the south/red.  The C region has a maximum observed angular extent of 1.37\arcsec\ from the center of the remnant, corresponding to 14.5~$\times$~10$^{17}$ cm and  16.7~$\times$~10$^{3}$ km s$^{-1}$ assuming free expansion.  


\begin{figure*}
\begin{center}
\epsfig{figure=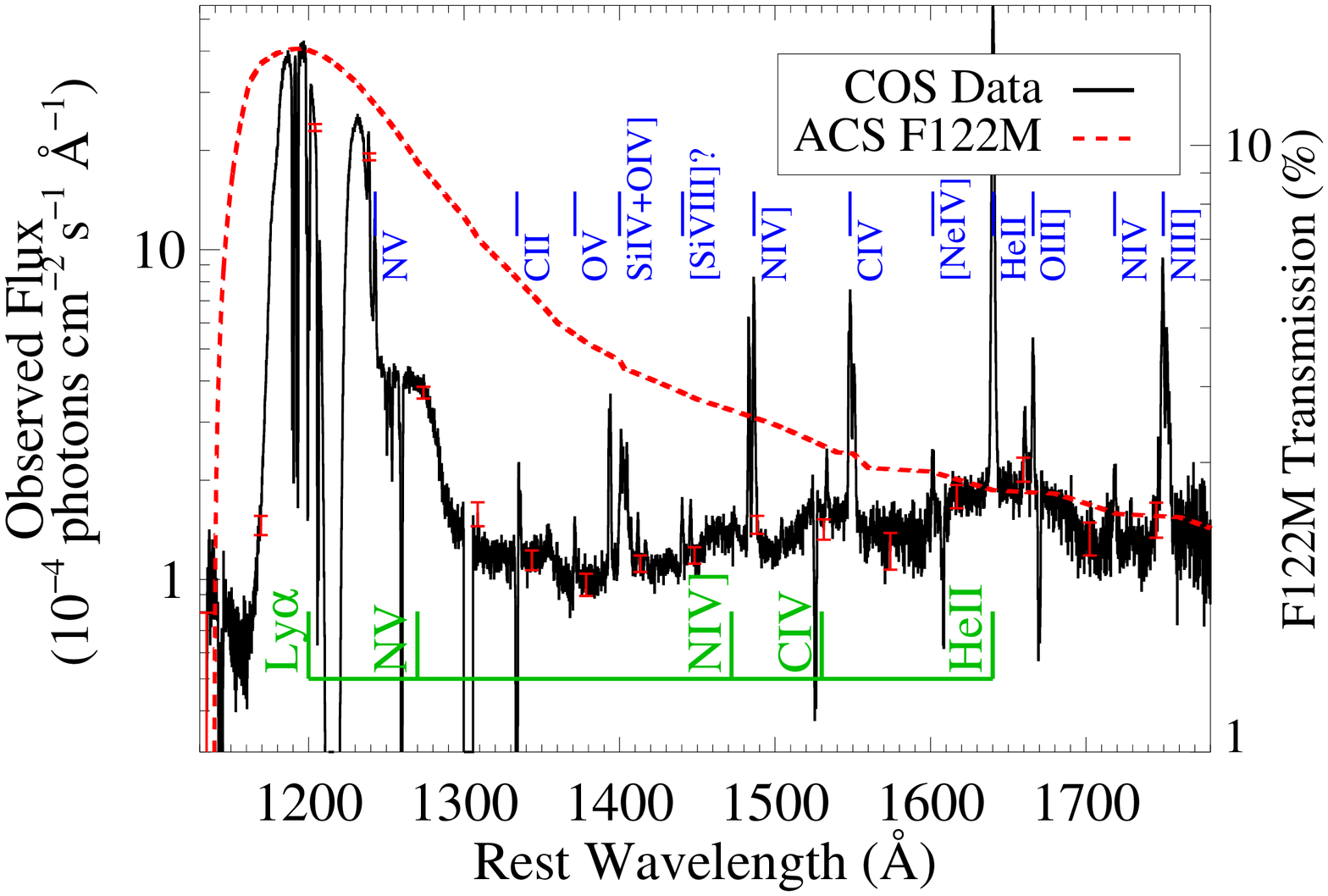,width=2.3in,angle=90}\hspace{+0.25in}
\epsfig{figure=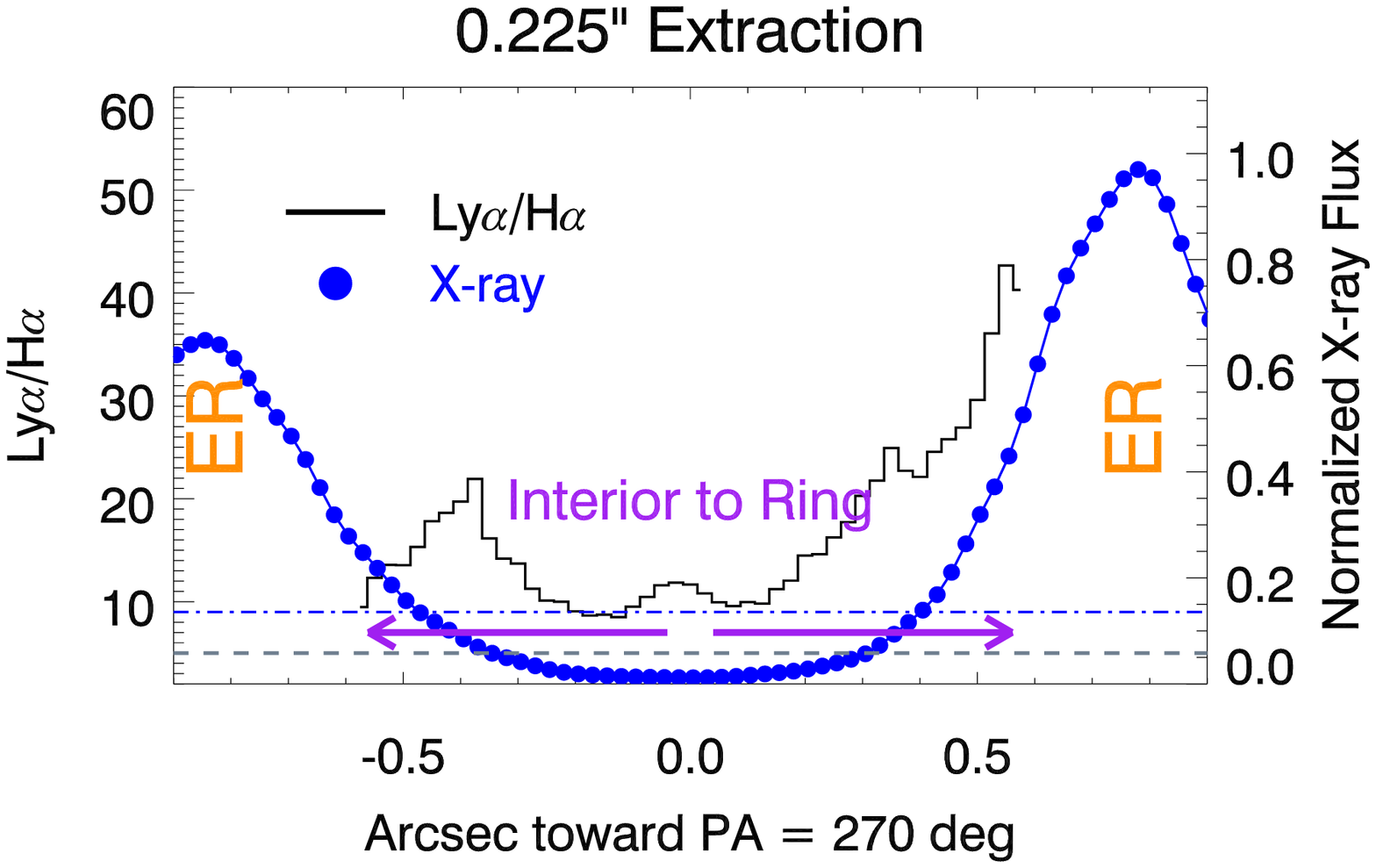,width=2.3in,angle=90}\vspace{-0.35in}
\epsfig{figure=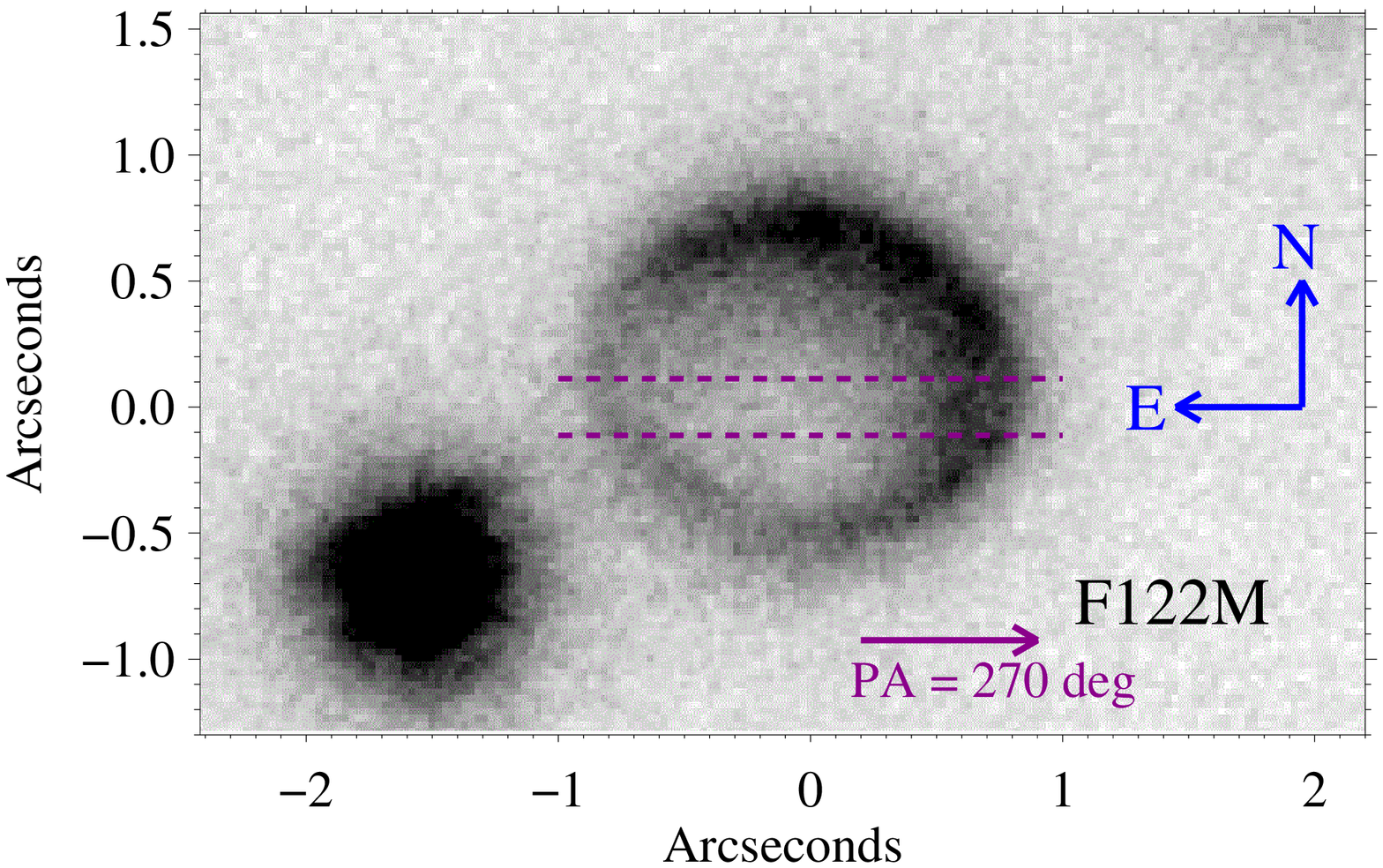,width=3.6in,angle=90}
\vspace{+0.05in}
\caption{
\label{cosovly} (top left) 2011 COS spectrum, with F122M filter curve overplotted (dashed red line). At upper right, $R(L\alpha / H\alpha)$ (photon flux ratio) in a 0.225\arcsec\ E-W spatial extraction region (orientation indicated on image below) with normalized X-ray flux in blue circles.  Emission from the ER prevents a measurement of $R(L\alpha / H\alpha)$ at angles $>$~$\pm$~0.6\arcsec.   $R(L\alpha / H\alpha)$ is elevated towards the regions of brightest X-ray emission. The ACS/SBC F122M Lyman-$\alpha$ image is shown at bottom (June 21 2014).}
\end{center}
\end{figure*}

\subsection{Morphology of the Reverse Shock Surface}

\citet{fransson13} noted that the large observed velocity of the H-$\alpha$ at late times ($V_{obs}$~$\gtrsim$~11~$\times$~10$^{3}$ km s$^{-1}$ at $t$~$>$~20 years) suggests that the reverse shock front has expanded beyond the ring plane.   They conclude that the RS extent in the polar direction is $\sim$~15\% larger than the ER radius and $\sim$~40\% larger than the RS radius in the ring plane.  We find de-projected velocities of 10~--~12~$\times$~10$^{3}$ km s$^{-1}$ in the B-b and B-r regions, corresponding to RS radii 40~--~70\% larger than the 6.1~$\times$~10$^{17}$ cm ring radius.  The C regions extend to 2~--~2.5 times the equatorial ring radius.  

To interpret the high-velocity H-$\alpha$ images, we recall that surfaces of constant Doppler shift are planar sections of the freely expanding supernova debris. Therefore, the pass bands of the ``H-$\alpha$ Blue'' and ``H-$\alpha$ Red'' filters map to slabs in physical space, as illustrated in Figure 5.   Portions of the RS surface visible through the ``H-$\alpha$ Blue'' filter (colored blue) reside within the slab delineated by the blue vertical lines, while those visible through the ``H-$\alpha$ Red'' filter (colored red) reside within the slab delineated by the red vertical lines.  The density of the outer debris crossing the RS front, and therefore the mass flux, is $\rho_{sn}$~$\propto$~$\rho_{o}$ $t^{-3}$ $V^{-9}$~\citep{luo94}.  The much higher density of the debris crossing the RS in the equatorial direction relative to the polar direction ($\rho_{ER}$/$\rho_{pol}$~$\sim$~10$^{4}$) makes the A region much brighter (detectable spectroscopically within $\sim$~10 years of the explosion).  The lower density material flowing out of the ring plane (B and C regions) is fainter, only now visible with the dedicated  high-velocity H-$\alpha$ imaging and sufficient time for the polar shock structure to be angularly separated from the ring plane.  

While much of the RS surface is not imaged because of the discrete velocity surfaces transmitted through the filters (shaded gray in Figure 5), the emerging picture for the extended H-$\alpha$ emission in the off-plane regions suggests a bipolar morphology.  This is expected due to the high pressure of the equatorial plane, owing to the presence of the ring~\citep{blondin96}.  Assuming the outer ejecta pressure is proportional to $\rho_{sn}$~$V^{2}$, the C polar region pressure is $\sim$~1/1800 of the equatorial region pressure.  This general picture resembles the bipolar surfaces suggested for the pre-supernova interacting wind picture by~\citet{blondin93}, 
however, 
recent studies of a `pre-explosion twin' to the SN 1987A system (SBW1; Smith et al. 2013) suggest that interacting winds may not be required to explain the nebular morphology.\nocite{smith13}

A second possible explanation for the C region emission is a conical or helical structure (e.g. Smith et al. 2007, 2013) connecting the ER with the extended outer rings (ORs).\nocite{smith07,smith13}  In this scenario, {\it 1)} high-velocity ejecta crossing this structure would be limb-brightened, explaining the bright edges in the C-b and C-r emission, and {\it 2)} the southern C-b streaks would be offset to the east compared to the northern C-r streaks by an amount comparable to the E-W offset of the ORs (e.g, Sugerman et al. 2005), which is indeed what is observed in Figure 2.  In this picture, the C emission would essentially be a new SN~1987A intermediate ring system. \nocite{sugerman05}
Deep H-$\alpha$ imaging spectroscopy to map the region could distinguish between these scenarios and may shed light on the progenitor structure, including the possibility of a binary merger prior to the supernova explosion~\citep{morris07}. \nocite{luo94,blondin96}

\begin{figure}
\begin{center}
\epsfig{figure=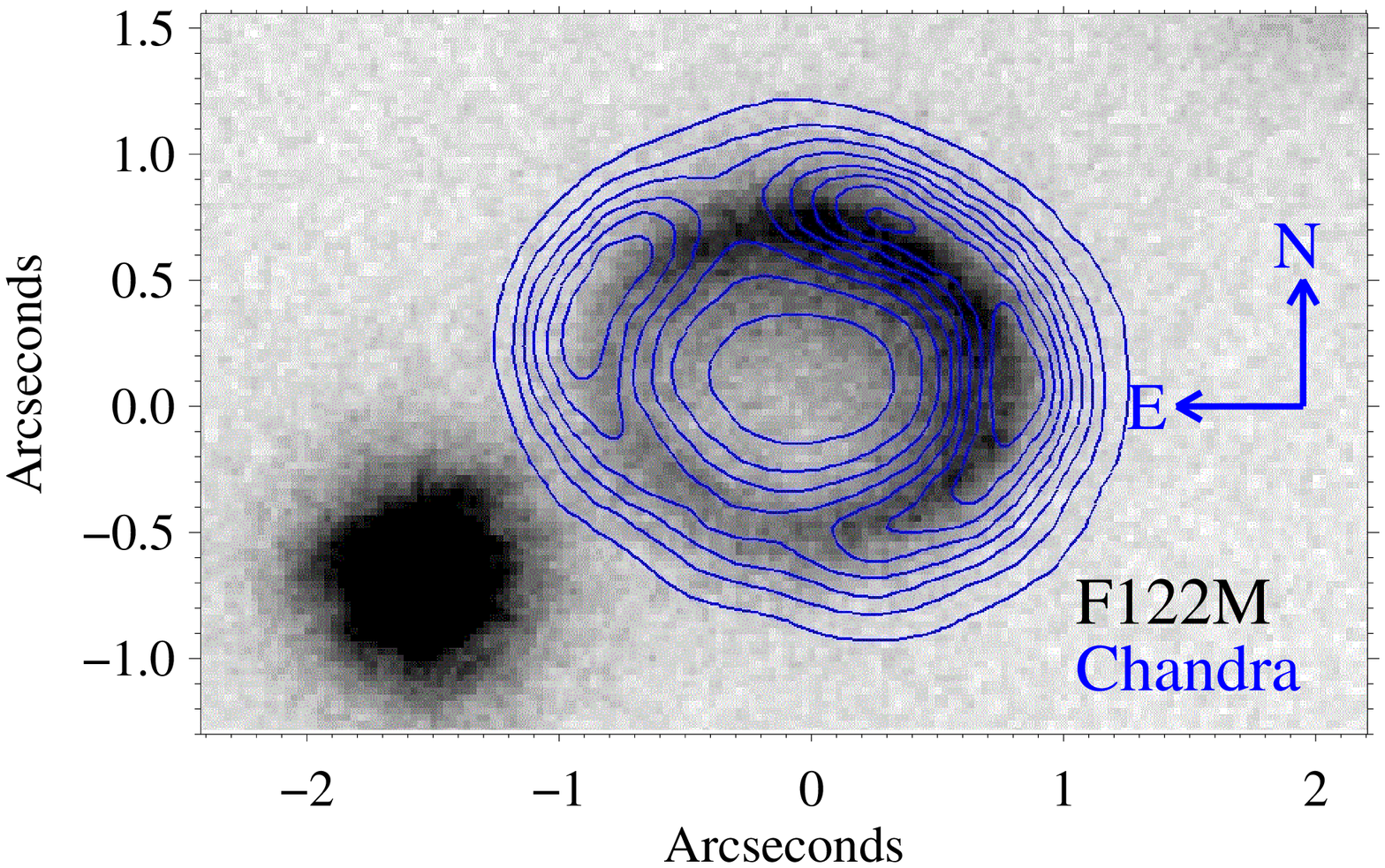,width=3.7in,angle=00}
\epsfig{figure=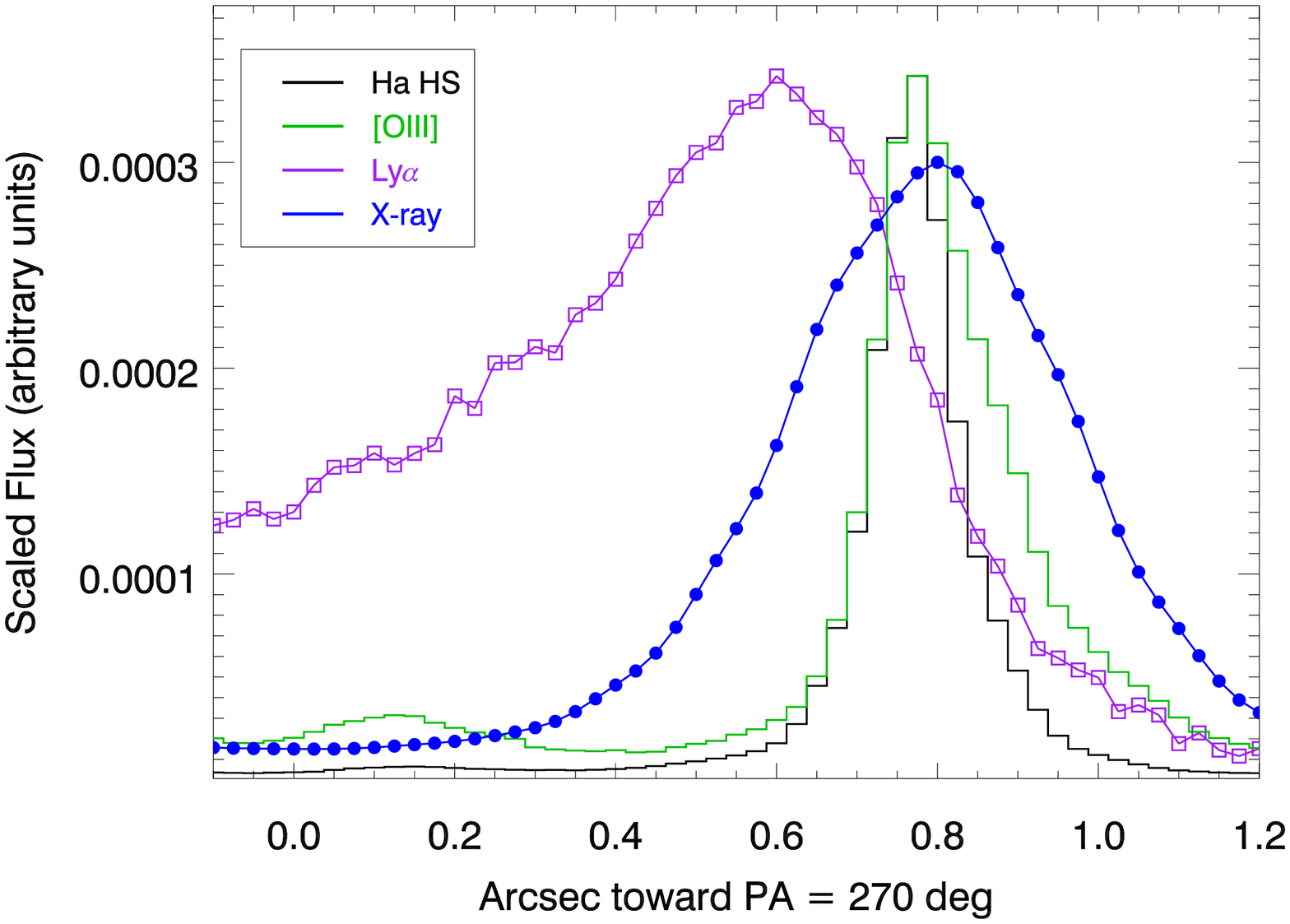,width=2.6in,angle=90}
\vspace{-0.25in}
\caption{
\label{cosovly} ACS/SBC F122M Lyman-$\alpha$ image with
19 March 2014 Chandra (0.3~--~8.0 keV) contours overlaid in blue (top).  Below, spatial profiles of the western edge of the ER region show that the Lyman-$\alpha$ emission arises interior to the equatorial ring as traced by X-ray (blue), [\ion{O}{3}] (green), and low-velocity H-$\alpha$ emission (black).  
 }
\end{center}
\end{figure}

\section{High-velocity Lyman-$\alpha$ Imaging}

 Figure 3 illustrates the contribution of different far-UV emission components to the F122M image.  Interstellar \ion{H}{1} blocks the center of the Lyman-$\alpha$ line profile over the velocity range $-$1500 km s$^{-1}$ $<$ $V_{obs}$ $<$ +1500 km s$^{-1}$~\citep{france11}, and so excludes emission from the shocked ring and hotspots, which have Doppler shifts $\Delta$ $V$ $<$ 300 km s$^{-1}$ (Pun et al 2002). Based on the $HST$-COS spectrum, we estimate that 90~--~95\% of the total emission in the F122M image is contributed by high-velocity Lyman-$\alpha$.  The remaining 5~--~10\% is mostly \ion{N}{5} $\lambda$~1240~\AA\ emission from the ER and high-velocity N$^{4+}$ ions crossing the RS front.  

Separation of the peak Lyman-$\alpha$ and ER emission is complicated on the bright northern side by the 43\arcdeg\ projection on the plane of the sky, however, there is separation between the Lyman-$\alpha$ and the H-$\alpha$ ring profiles in the East-West direction.  The East-West ring diameters are 1.44\arcsec\ ($\pm$~0.05\arcsec), 1.62\arcsec\ ($\pm$~0.05\arcsec), and 1.63\arcsec\ ($\pm$~0.10\arcsec) for Lyman-$\alpha$, H-$\alpha$ ER (F656N), and X-ray images, respectively.  The Lyman-$\alpha$ image is slightly offset with respect to the ER, with the largest separation (1.35~$\times$~10$^{17}$ cm) at the western edge (Figure 4, bottom).  
While the X-ray resolution is not as high as the $HST$ imaging of the ER, there is a clear offset between the X-ray and Lyman-$\alpha$ profile peaks.    Unlike the X-ray and Lyman-$\alpha$ images, which are brightest in the NW quadrant, the radio image~\citep{zanardo14} is brightest in the East.  

\begin{figure}
\begin{center}
\epsfig{figure=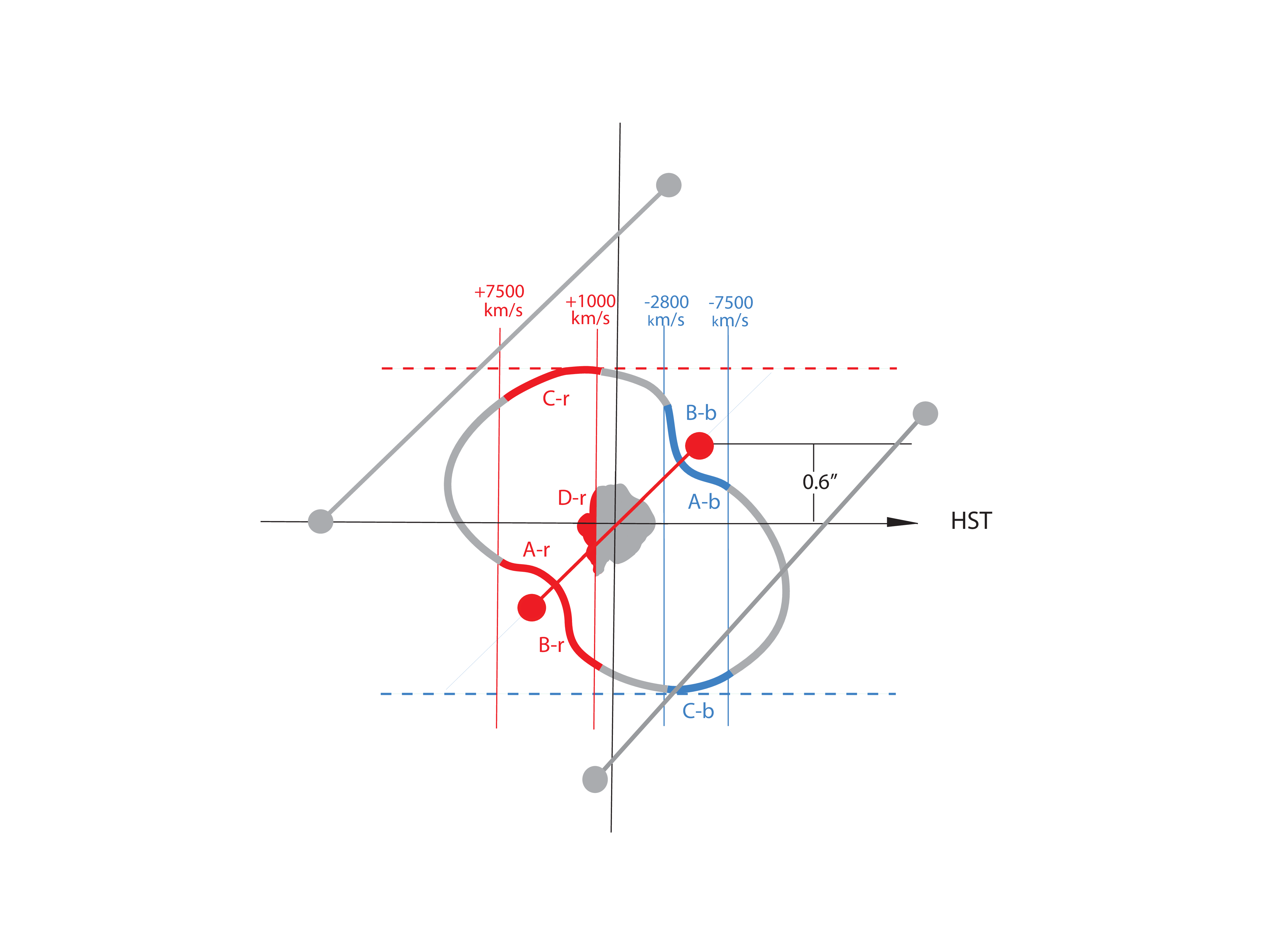,width=3.7in,angle=0} 
\vspace{+0.0in}
\caption{
\label{cosovly} Schematic representation of the underlying structure in the H-$\alpha$ imaging.  
The diagonal lines connecting the filled circles represent the circumstellar rings.  The red (blue) vertical lines delineate the slabs in which H-$\alpha$ emission from the reverse shock is visible through the ``H-$\alpha$ Red'' ( ``H-$\alpha$ Blue'') filter. 
The A, B, C, D labels refer to the image locations in Figure 2; the ``b'' and ``r'' refer to the ``H-$\alpha$ Blue'' and ``H-$\alpha$ Red'' images, respectively.  Regions colored blue and red appear in Figure 2, while regions colored gray are invisible to the H-$\alpha$ imaging.  
 }
\end{center}
\end{figure}

\subsection{The Lyman-$\alpha$/H-$\alpha$ Ratio: X-ray Heating}

{\it Neutral Hydrogen Mass Flux and the Lyman-$\alpha$/H-$\alpha$ Ratio~--~} If all of the high-velocity hydrogen emission originates from collisional excitation as neutral hydrogen atoms cross the RS front, the total flux is proportional to the fluence of hydrogen atoms across the shock boundary.  Integrating the individual band images, we calculate the total number of atoms in a given velocity range.  We take the total (Galactic + LMC) reddening towards SN 1987A as a standard ISM curve with $R_{V}$ = 3.1 and $E(B~-~V)$~=~0.19 (see France et al. 2011 for a discussion).  The total rate of hydrogen atoms crossing the RS front is approximately:
\begin{equation}
\dot{N_{H}}~=~4\pi d^{2} \times F(\Delta\lambda)  \frac{\lambda}{hc} \Delta\lambda \times P(H)^{-1} \times C_{ISM}
\end{equation}
where $d$~$\sim$~50 kpc ($\sim$~10\% uncertainty; Panagia et al. 1991), $F(\Delta\lambda)$ is the image-integrated flux density in a given band ($\sim$~5\% uncertainty), $\Delta\lambda$ is the effective bandpass of the filter, $P(H)$ is the number of line photons emitted per atom, and $C_{ISM}$ is the interstellar attenuation correction factor.   $P(H)$ $\approx$~0.2 for H-$\alpha$ and 1 for Lyman-$\alpha$~\citep{michael03,heng07}.

For the ``H-$\alpha$ Blue'' image, 
 $\Delta\lambda$~$\approx$~100~\AA\ and $C_{ISM}$ = 1.57~$\pm$~0.08, giving a total hydrogen fluence of 4.1~($\pm$~0.7)~$\times$~10$^{46}$ atoms s$^{-1}$.  For the ``H-$\alpha$ Red'' image, 
$\Delta\lambda$~$\approx$~120~\AA\ and $C_{ISM}$ = 1.53~$\pm$~0.08, giving a total hydrogen fluence of 3.3~($\pm$~0.5)~$\times$~10$^{45}$ atoms s$^{-1}$.  
Taken together, we estimate the total mass flux of high-velocity hydrogen atoms crossing the RS is $\dot{M_{H}}$ = 7.4~($\pm$~1.2)~$\times$~10$^{22}$ g s$^{-1}$ (1.2~$\times$~10$^{-3}$ M$_{\odot}$ yr$^{-1}$).  

The analogous calculation using the F122M Lyman-$\alpha$ image has $\Delta\lambda$~$\approx$~100~\AA\ and $C_{ISM}$ = 6.7~$\times$~1.63, where 6.7~($\pm$~1.7) is the correction for dust attenuation and 1.63~($\pm$~0.08) is the empirically determined correction for the interstellar Lyman-$\alpha$ line core attenuation. 
This calculation 
shows a Lyman-$\alpha$-to-H-$\alpha$ photon ratio, $R(L\alpha / H\alpha)$, of $\approx$~17~$\pm$~6,
a factor of 3.5 larger than attributable to excitation in the RS alone.   Therefore, an additional Lyman-$\alpha$ emission mechanism is present.  Note that the reddening correction does not account for differential extinction within the supernova debris, so the intrinsic $R(L\alpha / H\alpha)$ may be greater.  Lyman-$\alpha$ excess was observed by previous authors using spectroscopic observations of small spatial regions of the SN 1987A inner ring region~\citep{heng06,france10c}.  

Combining the ``H-$\alpha$ Blue'' and ``H-$\alpha$ Red'' images with the Lyman-$\alpha$ map, we find the baseline $R(L\alpha / H\alpha)$ interior to the ER is 8~--~10, at the center and southeastern sides of the inner remnant.  $R(L\alpha / H\alpha)$ reaches a maximum on the western and northwestern side of the inner remnant where the $Chandra$ emission is brightest (Figure 3).  The maximum $R(L\alpha / H\alpha)$ is $\sim$~35 near PA = 270\arcdeg, just interior to the brightest X-ray emission from the ER.  $R(L\alpha / H\alpha)$ values $\sim$~20 are found towards PA~$\sim$~60\arcdeg, the location of the northeast X-ray maximum.  

{\it The Role of X-rays~--~}  The observed offset between the X-ray and Lyman-$\alpha$ emitting regions is consistent with the expected separation between the ER and the outer extent of the RS.   Following the observation that the dominant energy input in the unshocked supernova debris has evolved from internal heating by radioactive decay to external heating from X-rays produced at the ER~\citep{larsson11}, \citet{fransson13} presented calculations of the X-ray energy deposition into the ejecta.   The heating of the outer debris is dominated by X-rays with $E$~$<$ 0.5 keV while higher energy X-rays penetrate deeper to heat the interior.  X-rays propagating into the outer debris produce fast photoelectrons, which deposit most of their kinetic  energy as heat, provided that the debris has fractional ionization $n_e/n_H$ $>$ 0.03~\citep{xu91}.  Radiative cooling by thermal excitation of Lyman-$\alpha$ limits the debris temperature to $T$~$\lesssim$~ 10$^{4}$ K, at which temperature thermal excitation of H-$\alpha$ is negligible (this mechanism, and not recombination, also accounts for the \ion{H}{1} two-photon spectrum observed in the UV, France et al. 2011).   Therefore, we expect the Lyman-$\alpha$/H-$\alpha$ ratio to be at a maximum nearest the source of the X-ray heating.  The spatial stratification of the Lyman-$\alpha$ and X-ray emission and the enhanced Lyman-$\alpha$/H-$\alpha$ ratio support the scenario where X-ray heating of the outer debris is responsible for the majority of the observed Lyman-$\alpha$ emission from SN 1987A at present.   

The total Lyman-$\alpha$ luminosity in the F122M band is $L_{Ly\alpha}$ = 2.5~($\pm$~0.7)~$\times$~10$^{36}$ erg s$^{-1}$, $\sim$~70\% of which is in excess of the expected RS emission.  The total flux of the X-ray ring is 1.04~$\times$~10$^{-11}$ erg cm$^{-2}$ s$^{-1}$, or, $L_{X}$(0.3~--~8.0 keV) = 3.1~($\pm$~0.3)~$\times$~10$^{36}$ erg s$^{-1}$.  However, most of the heating radiation is in the soft X-ray/EUV band (0.01~--~0.5 keV) which is not observed at Earth because of interstellar neutral hydrogen and dust attenuation towards SN 1987A.  Scaling the Zhekov et al. (2006) model X-ray spectrum to the observed day 9885 0.3~--~8.0 keV flux, the total 0.01~--~8.0 keV luminosity is $L_{X}$(0.01~--~8.0 keV) = 5.6~$\times$~10$^{36}$ erg s$^{-1}$.\nocite{zhekov06}  Assuming that the only source of EUV/soft X-rays is that observed by $Chandra$ and that the neutral hydrogen in the outer ejecta intercepts half of the soft X-rays emitted by the ring, the Lyman-$\alpha$ heating efficiency must be $\sim$~60\% to reproduce the F122M image of SN 1987A.  However, slower shocks that contribute more to the optical/UV emission of the ER than to the X-rays will also be a source of EUV irradiance of the outer debris~\citep{fransson13}, meaning that the actual soft X-ray heating efficiency is likely substantially lower.  

 
The $HST$ observations presented here were acquired as part of the Cycle SAINTS program (13401 and 13405).

\acknowledgments
The $HST$ observations presented here were acquired as part of the Cycle SAINTS program (13401 and 13405).  




\end{document}